\documentclass[smallextended]{svjour3}       % onecolumn (second format)
\smartqed  % flush right qed marks, e.g. at end of proof
\usepackage{graphicx}
\usepackage{amsmath}
\usepackage{amsfonts}
\usepackage{amssymb}
\usepackage{epstopdf}
\usepackage{subfig}

\newcommand{\echo}{EChO}
\newcommand{\echosim}{EChOSim}
\newcommand{\echodp}{EChOSim-DP}

\begin{document}

\title{Data analysis Pipeline for EChO end-to-end simulations%\thanks{Grants or other notes
}


\author{Ingo P. Waldmann \and E. Pascale}


\institute{Ingo P. Waldmann \at
              Dept. Physics \& Astronomy, University College London, Gower Street, London, WC1E 6BT, UK \\
              \email{ingo@star.ucl.ac.uk}           %  \\
           \and
           E. Pascale \at
              School of Physics \& Astronomy, Cardiff University, Cardiff, CF24 3AA, UK \\
}

\date{Received: date / Accepted: date}

\maketitle

\begin{abstract}
Atmospheric spectroscopy of extrasolar planets is an intricate business. Atmospheric signatures typically require a photometric precision of $1 \times 10^{-4}$ in flux over several hours. Such precision demands high instrument stability as well as an understanding of stellar variability and an optimal data reduction and removal of systematic noise. In the context of the \echo~mission concept, we here discuss the  data reduction and analysis pipeline developed for the \echo~end-to-end simulator \echosim. We present and discuss the step by step procedures required in order to obtain the final exoplanetary spectrum from the \echosim~`raw data' using a simulated observation of the secondary eclipse of the hot-Neptune 55 Cnc e.

\end{abstract}

\section{Introduction}
\label{intro}

The field of extrasolar planets is innovative as it is new. Recent successes in characterisation of extrasolar planets are also always tales of characterising the instrument response function to an unprecedented detail. Always being at the edge of technical feasibility means that instrument calibration, observing strategy as well as data analysis and modelling are interdependent.  In the light of the \echo~ESA-M3 mission concept \cite{tinetti12}, such interdependence becomes important in the study of engineering decisions and instrument trade-offs. In other words, one needs to simulate the full observational and data analysis chain in order to gauge the impact the instrument concept has on the achievable error bar of the detection. Such a feat requires an advanced mission end-to-end simulator as well as an advanced data analysis pipeline. In this paper, we discuss the data analysis pipeline which is used in conjunction to the  mission simulator, \echosim~ \cite{pascale13}. The \echosim~data pipeline (from here on \echodp) is a stand-alone software custom built for \echosim~but with easy adaptability to other instruments and data sets in mind.

The method by which the \echo~mission will characterise the nature of extrasolar planets is by time resolved spectroscopy of their atmospheres, in particular of transiting extrasolar planets. Briefly, when an exoplanet transits in front of its host star (in our line of sight) we observe a diminishing of the stellar flux due to the obscuration of the planet. The depth of the resulting lightcurve allows us to estimate the planetary radius (given the stellar radius is known). This we refer to as transit (or primary eclipse) observation. Should the exoplanet feature an extended atmosphere, we expect some of the stellar light to filter through the terminator region of the planetary atmosphere. Here we are sensitive to molecules absorbing the stellar light at specific wavelengths. We hence perceive a variation of transit depths depending on the wavelength range observed. These variations constitute the signatures of an exoplanetary absorption spectrum. Similarly, we can observe the occultation (or secondary eclipse) where the thermal contribution of the exoplanet's day-side is lost to the observer as the planet passes behind its host star. 
The study of transmission and emission spectroscopy is now a well established field for both space and ground based observations of exoplanetary atmospheres (e.g.
\cite{beaulieu10,beaulieu11,charbonneau08,brogi12,bean11,swain08,swain08b,swain09a,crouzet12,deming13,grillmair08,thatte10,tinetti07,pont08,swain12,knutson11,sing11,tinetti10,mooij12,bean11b,stevenson10} also see \cite{tinetti13} for a comprehensive review). 

\subsection{\echosim}

\echosim~ is the \echo~ mission end-to-end simulator. \echosim~  implements a detailed simulation of the major observational and instrumental effects, and associated systematics. It also allows the influence of individual instrumental and astrophysical parameters to be studied and thus represents a key tool in the optimisation of the instrument design. Observation and calibration strategies, data reduction pipelines and analysis tools can all be designed effectively using the realistic outputs produced by \echosim \cite{pascale13,waldmann13}.
The simulation output closely mimics standard STSci\footnote{$http://archive.stsci.edu/hst/$} FITS files, allowing for a high degree of compatibility with standard astronomical data reduction routines. 

\subsection{Examples}

We illustrate individual steps in \echodp~using diagrams. Unless specified otherwise, we follow a single data processing run of \echosim~simulated data of the hot-Neptune 55 Cnc e. \echosim~was run to simulate the Chemical Census mode of \echo, in which we co-add (in the case of 55 Cnc e) five eclipse observations to obtain a minimal signal-to-noise  (S/N) of the final spectrum of S/N $\sim$ 5. 

For this we assume spectra reconstructed with a resolving powers of 50,50,30,30,30 for the VNIR, SWIR, MWIR-1, MWIR-2, and LWIR channels. With each channel having a larger native resolving power, this allows us to increase the SNR of the detection for this particular observing mode. See \cite{tinetti12} for a review of the proposed \echo~observing modes.

%For this we assume a strong spectral binning (i.e. low spectral resolution) of  R = 50, 50, 30, 30, 30 for the five detectors of \echo~ranging from visible to far-IR respectively.  

\section{Data Reduction}

\begin{figure}
  \centering
    \includegraphics[width=0.5\textwidth]{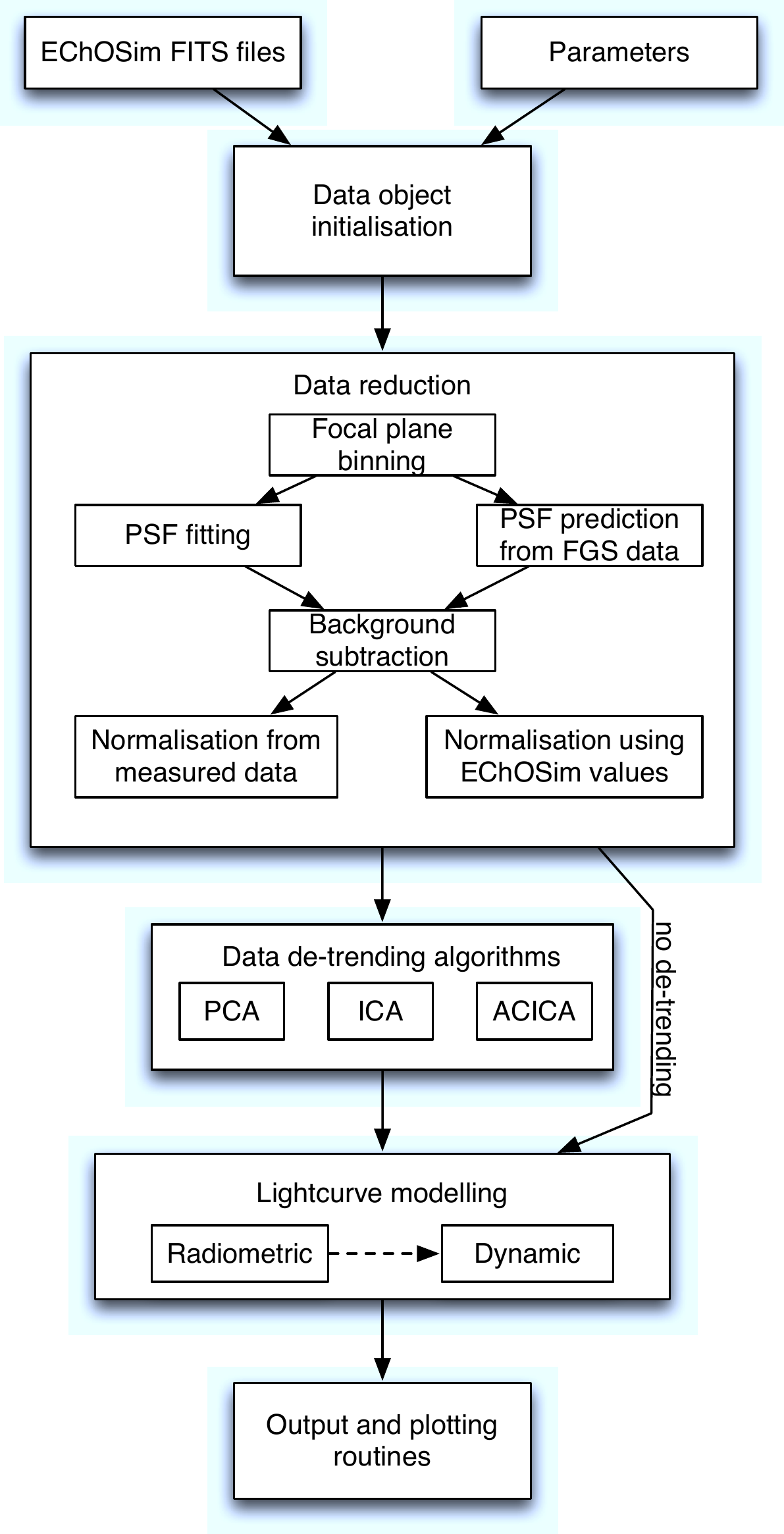}
   \caption{Flowchart of the \echodp~design. The pipeline is subdivided into five main modules (contained as individual python classes): 1) Object initialisation and data read, collating all input data and parameter files and performing format conversions where necessary, 2) Data reduction, reducing the two dimensional focal plane images to 1D wavelength dependent time series, 3) de-trending all or individual time-series using non-parametric machine learning techniques, 4) model fitting the final lightcurve, 5) collecting all data and calculating the final spectrum.  \label{fig:flowchart}}
   
\end{figure}

The \echodp~is a stand-alone package delivered with the \echosim~code but can easily be adapted to observations produced by any spectrograph. It is written in fully object orientated Python allowing for a cross platform compatibility and an easy adaptability through its modular design. \echodp~is subdivided into five main modules: 1) The data and parameter read-in and object initialisation, 2)  data reduction, going from two dimensional focal plane illuminations to 1D time series data, 3) time series de-trending using non-parameteric de-trending algorithms, 4) lightcurve fitting using simplex-downhill minimisations as well as Markov Chain Monte Carlo (MCMC) techniques, 5) collection of results and computation of the final spectrum. We summarise this flow in figure~\ref{fig:flowchart}.

\subsection{Configuration and data formats}

The output of \echosim~follows the standard FITS file conventions with the aim to make the raw data generated by \echosim~as universally readable as possible. The payload of \echo~is subdivided into individual channels defined as: VNIR (0.4 - 2.5$\mu$m), SWIR (2.5 - 5.0$\mu$m), MWIR-1 (5.0 - 8.5$\mu$m), MWIR-2 (8.5 - 11.0$\mu$m) and LWIR (11.0 - 16.0$\mu$m). For a detailed description of the individual channels we refer the reader to \cite{tinetti12} and publications in this special issue.
Due to varying detector array sizes, it is not possible to combine all focal plane read-outs (for an individual frame) in one conventional FITS data-cube. \echosim~hence utilises extensions to the Primary FITS Header Data Unit (PrimaryHDU). This allows the inclusion of meta data on each detector as well as additional auxiliary information carried in binary tables (BinaryHDUs). \echosim~produces one FITS file per integration interval resulting in 10s to 100s of files per simulated observation run. Whilst the high number of output files produced seems cumbersome, it reflects the data handling strategies of current space and ground based instruments. 
\echodp~is designed to be fully compatible to this customised FITS convention using a custom build read-in routine based on the PyFITS\footnote{$http://www.stsci.edu/institute/software_hardware/pyfits$} package. \echodp~can also natively read single HDU FITS files generated by other instruments. 

Auxiliary information contained in BinaryHDUs contains: the \echosim~generated stellar limb-darkening grid, \echosim~generated noiseless stellar, zodi and thermal fluxes from the instrument and its optical elements, \echosim~generated exoplanetary eclipse/transit depths, \echosim~generated Keplarian solutions. If specified by the user, \echodp~can use these auxiliary data to calculate exact time series normalisation constants and eclipse/transit models to estimate best-case scenarios.

\echodp~specific parameters are specified in a separate ascii file and parsed using the python specific ConfigParser\footnote{http://docs.python.org/2/library/configparser.html}. 
%An example of the parameter files can be found in appendix XXX.

\begin{figure}
  \centering
    \includegraphics[width=0.85\textwidth]{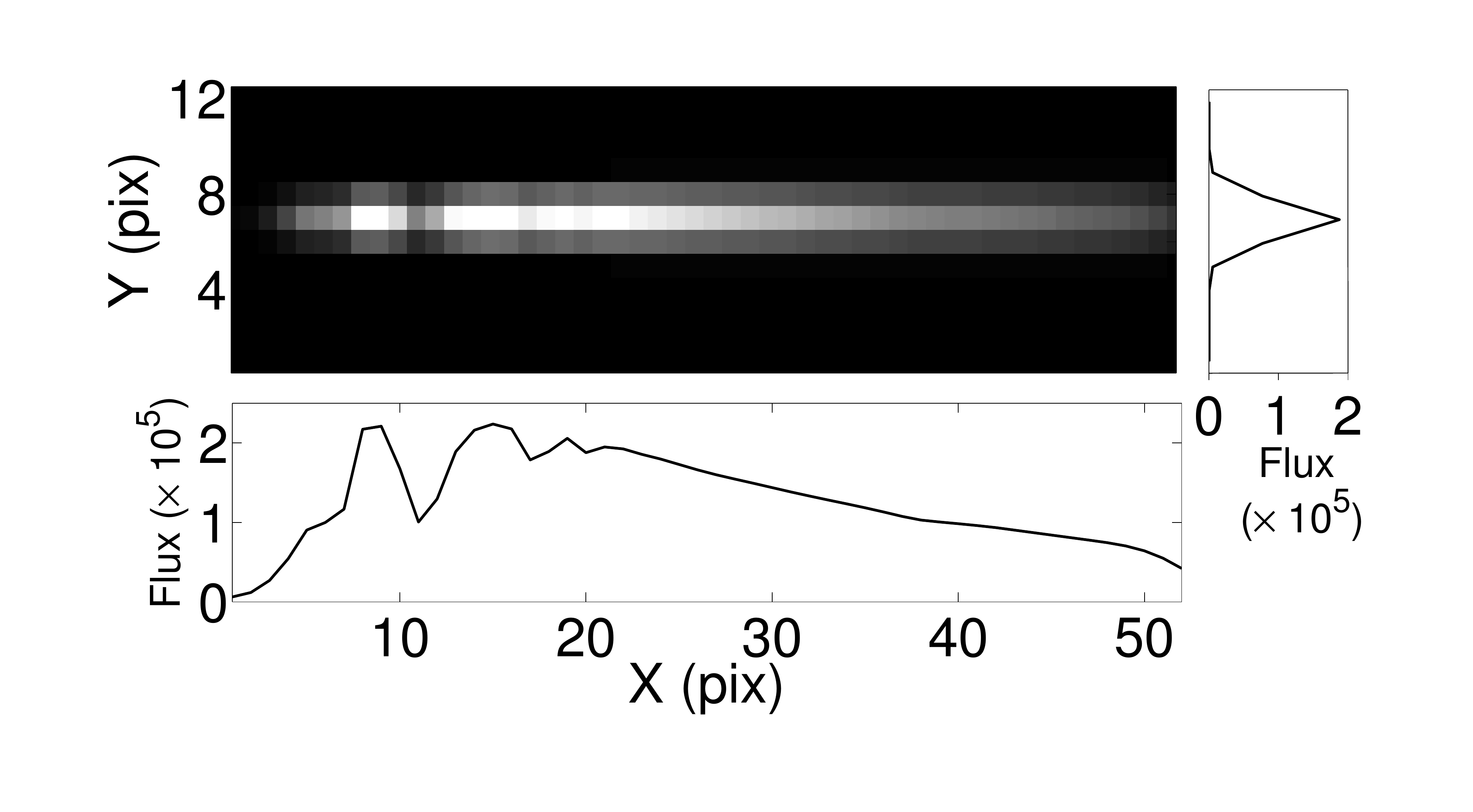}
   \caption{Top left: Focal plane of the mid-IR2 detector as read in by \echodp. Bottom and right: cross cuts though the focal plane along the spectral and spatial directions respectively. Most flux is contained within three pixels of the spatial direction. \label{fig:focalplane}}
\end{figure}

\subsection{Focal Plane Binning}

\begin{figure}
  \centering
    \includegraphics[width=0.85\textwidth]{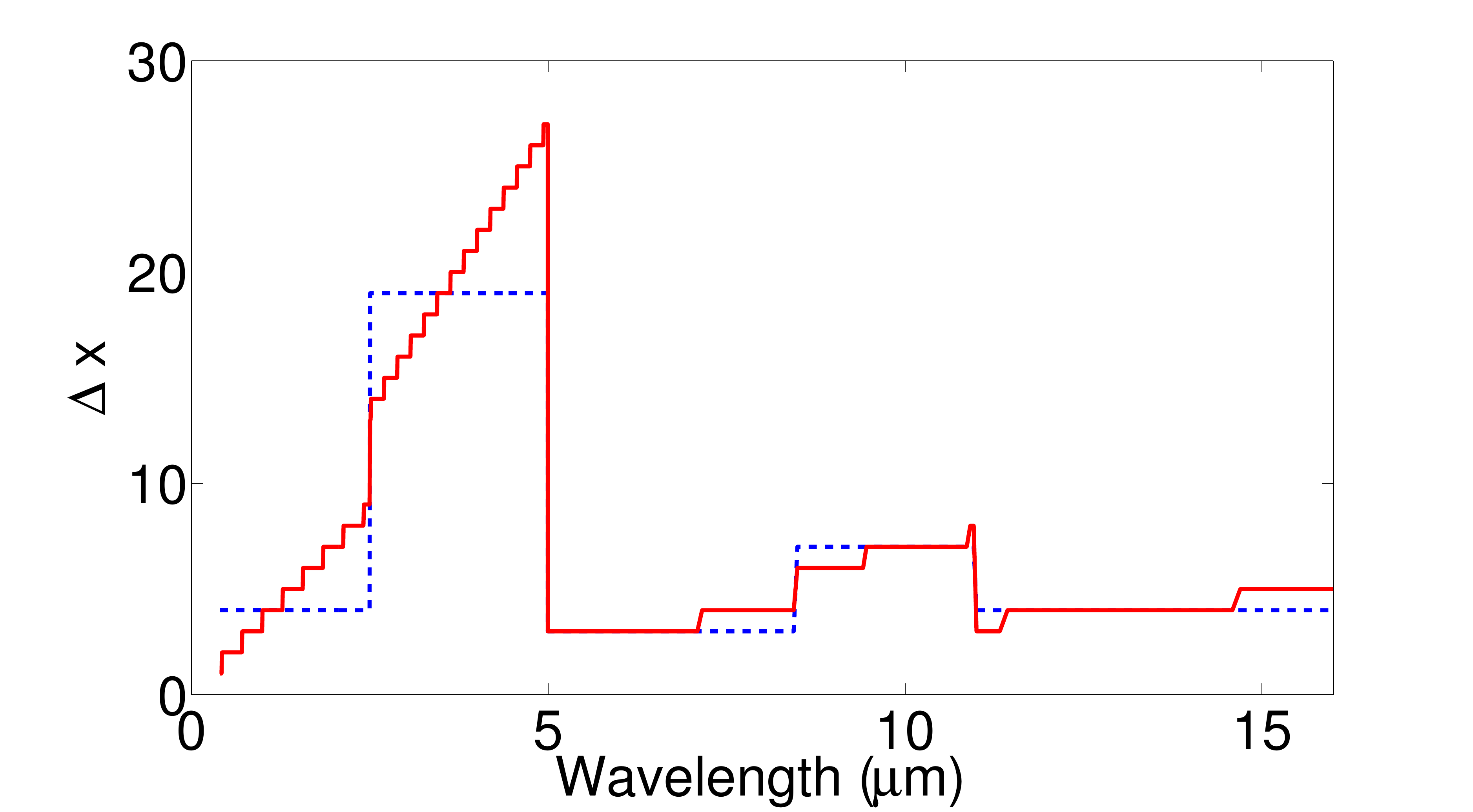}
   \caption{showing binning steps $\Delta x$ in pixels as function of wavelength for the two spectral binning modes available in \echodp. Red-solid line shows the constant in R binning; blue-discontinuous line shows constant in $\Delta \lambda$ binning. A binning of R= 50,50,30,30,30 for all five detectors going from VIS to FIR respectively is assumed. \label{fig:bingrid}}
\end{figure}

Given current detector design specifications, the native spectral resolution ($R = \lambda /\Delta \lambda$) of \echo~can exceed that required by the science case.
\echodp~provides two available spectral binning formats: 1) constant R, equation~\ref{eq:constR}; 2) constant $\Delta \lambda$, equation~\ref{eq:constlambda}:

%\begin{equation}
%LD = \frac{\Delta x}{\Delta \lambda} = \frac{2 \Delta_{pix}}{\lambda} R(\lambda)
%\end{equation}

\begin{align}
\label{eq:constR}
\Delta x &= \frac{\lambda R(\lambda_{mid}) }{2\Delta_{pix}}\\
\Delta x&= \frac{\lambda_{mid} R(\lambda_{mid}) }{2\Delta_{pix}}
\label{eq:constlambda}
\end{align}

\noindent where $\Delta x$ is the binning interval along the spectral axis in pixels, $\lambda$ and $\lambda_{mid}$ the wavelength and central wavelength in $\mu$m and  $\Delta_{pix}$ the pixel size in $\mu$m. Note that \echo~ spectrometers sample each spectral resolving element with two detector pixels. Figure~\ref{fig:bingrid} shows $\Delta x$ for both binning methods as function of $\lambda$. Binning is performed directly on the focal plane before spectral extraction. This increases S/N and avoids potential biasing of the data. 

\subsection{Optimal extraction}
\label{sec:extraction}

\begin{figure}
  \centering
    \includegraphics[width=0.85\textwidth]{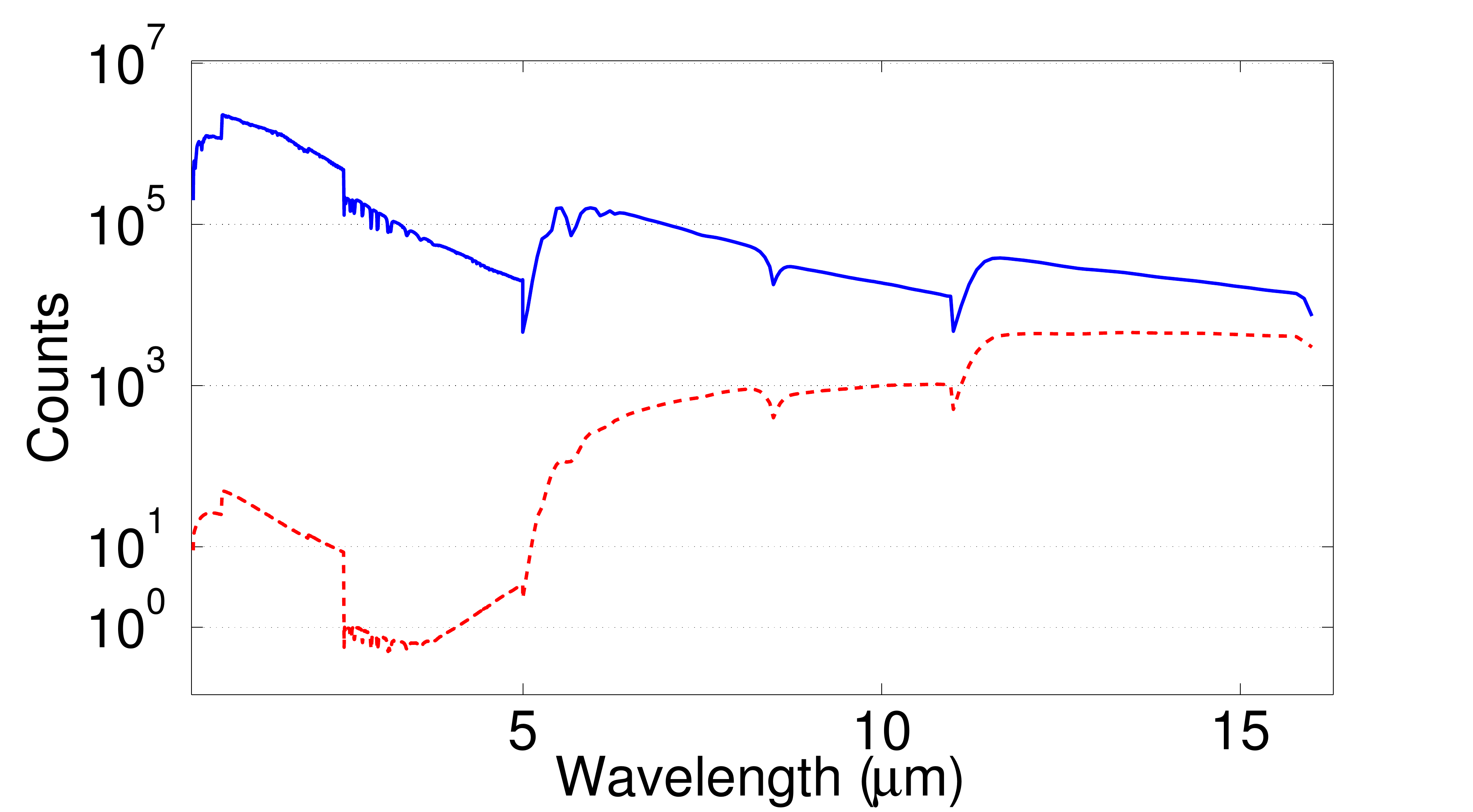}
   \caption{showing the extracted flux for a single frame as function of wavelength. Blue-continuous line: Optimally extracted flux before background subtraction; red-discontinuous line: estimated background counts measured on the off-axis spatial direction.  \label{fig:fluxes}}
\end{figure}

%I would change this by saying
%1) At each intergration time, the raw spectrum is extracted from the data by fitting a model of the PSF to the point-like dispersed signal of the star + planet. 
%2) The PSF is models can be chosen from  either Gaussian or a Voigt profile convolved with the detector response discussed in barrow et al. 
%The user decides the mode the PSF is fit to the data. 
%a) The width, position and amplitude are left to vary. 
%b) Only the amplitude is fit to data. In this case the width and shape of the PSF is constrained from instrument modelling or from an empirically determined PSF. The position of the PSF on the focal plane is constrained by the housekeeping information provided by the onboard fine guidance sensor, which monitors the location of the parent star in the FOV of the instrument.

After the data has been binned, we extract the raw spectrum along the spatial axis for each individual time stamp. At each integration time, the raw spectrum is extracted from the data by fitting a model of the PSF to the point-like dispersed signal of the star + planet flux.
Two extraction options are available: 1) Unconstraint PSF, 2) EChOSim PSF with Fine Guidance Sensor (FGS) offset data.

\emph{Option 1:} is the least constraint extraction. Depending on user input, \echodp~fits a Gaussian or Generalised Gaussian Distribution (GGD) PSF along the spatial axis. The GGD is given by

%%\begin{subequation}
%\begin{align}
%\label{eq:psfgauss}
%&PSF_{gauss}(y, t; \mu_{y}, \sigma_{y},\Delta y) = \\\nonumber
%& =  \frac{1}{\sigma_{y}  \sqrt{2 \pi} } \text{exp} \left (-\frac{(\mu_{y} + \Delta y(t)) - y}{2\sigma_{y}} \right )
%\end{align}
%%\end{subequation}

%\begin{align}
%\label{eq:psfvoigt}
%&PSF_{voigt}(y; \sigma_{y}, \gamma, \mu_{G}, \mu_{L},\Delta y) = \\\nonumber
%&= e^{{}i (\mu_{G} + \mu_{L}+\Delta y(t))y - \frac{\sigma^{2} y^{2}}{2} - \gamma |y|}
%\end{align}

\begin{align}
\label{eq:psfggd}
PSF_{ggd} = \frac{\beta}{2 \alpha \Gamma(1/\beta)} \text{exp} - \left [|(\mu_{y} + \Delta y(t)) - y|/\alpha \right ]^{\beta}
\end{align}

\noindent where $\mu_{y}$ is the mean position of the spectrum along the spatial axis $y$ for all frames, $\Delta y(t)$ is a time dependent offset from the mean, $\alpha$ is a scale parameter and in this case equivalent to $\alpha = 2 \sigma_{y}$ and $\sigma_{y}$ signifies the width of the PSF. The shape parameter $\beta$ introduces a kurtosis argument in the Gaussian distribution.  We retrieve the Normal PSF by setting $\beta = 2$ and obtain leptokurtic and platiokurtic distributions for $\beta < 2$ and $\beta > 2$ respectively. We do not assume skew of the PSF in the spatial direction. The PSF shape can either be left as free parameter (to be fitted from the data) or specified as user input.
Equation \ref{eq:psfggd} is convolved with the detector response function assumed by \echosim~to obtain the extraction profile.

\begin{equation}
\mathcal{P}(y,t) = PSF(y,t) \otimes \mathcal{R}(y)
\end{equation}

\noindent where $\otimes$ is the convolution operator and the detector response \cite{barron07} is given by 

%$\sigma_{y} = K_{y}\lambda$
%
%\noindent where $\mu_{y}$ is the mean position of the spectrum along the spatial axis $y$, $K_{y}$ is the aberration parameter, $\lambda$ is the wavelength in $\mu$m, $\Delta y (t)$ the offset from the mean due to photometric jitter for frame (or time interval) $t$. We convolve this idealised PSF with the instrument response function which is assumed to be 

\begin{align}
\label{eq:response}
&\mathcal{R}(y; \Delta_{pix},l_{y}) = \\\nonumber
& = \frac{\text{tan}^{-1} \left (\text{tanh}( \frac{\Delta_{pix}-y}{4l_{y}}) \right ) - \text{tan}^{-1} \left (\text{tanh}( -\frac{\Delta_{pix}-y}{4l_{y}}) \right ) }{\text{tan}^{-1} \left (\text{tanh}( \frac{\Delta_{pix}}{4l_{y}}) \right ) - \text{tan}^{-1} \left (\text{tanh}( -\frac{\Delta_{pix}}{4l_{y}}) \right )}
\end{align}

\noindent where $\Delta_{pix}$ is the pixel size in $\mu$m and $l_{y}$ the diffusion length in $\mu$m. 

\emph{Option 2:} Here we assume a Gaussian PSF (by setting $\beta = 2$) with a fixed width given by $\sigma_{y} = F\# K_{y}\lambda$ where $F\#$ is the effective focal length of the telescope in $\mu$m, $K_{y}$ is the PSF aberration parameter and $\lambda$ the wavelength in $\mu$m. 
We hold $\mu_{y}$ fixed at an \echosim~specified value and obtain the time dependent offset $\Delta y(t)$ from the \echosim~provided fine guidance sensor (FGS) centroiding.  

The centroiding is provided as part of the auxiliary information BinaryHDUs and consists of a time series of y-positional offsets sampled at 1Hz frequency. \echodp~downsamples the positional offsets to the integration times specified in the FITS headers. The downsampling operation correctly reflects the error in the positional offset $\Delta y (t)$ and the associated flux error.

\echodp~calculates the background by computing the median (or mean given user input) focal plane illumination $4 \sigma_{y}$ away from $\mu_{y}$. The background flux is integrated over the area (in pixels) of the extraction profile and subtracted form the extracted flux. 

\subsection{Normalisation}
\label{sec:extraction}

The final step is the normalisation of the data to the out of transit (OOT) baseline. Similarly to section~\ref{sec:extraction} the normalisation can either be estimated from the data itself by calculating the OOT mean or normalised using noiseless stellar fluxes provided by \echosim

\begin{equation}
F_{norm}(\lambda,t) = \frac{F_{total}(\lambda,t)}{F_{star}(\lambda,t)}
\end{equation}

\section{Data de-trending}

After the data as been reduced to 1D time series, \echodp~can attempt a de-correlation of wavelength correlated non-Gaussian systematics. These systematics tend to be due to array wide fluctuations of quantum efficiencies, insufficient flat-fielding, slit-loss effects and pointing jitter. These complex non-Gaussian signals have shown to be important effects in real instruments \cite{thatte10,waldmann12,waldmann13b,waldmann13c}. \echosim implements inter and intra-pixel variations and non-Gaussian pointing jitter noise. Other non-linear noise sources such as correlated astrophysical noise (e.g. such as stellar pulsation, stellar spots and faculae noise) will be included in future releases of \echosim. 

Here we implement the ACICA de-trending algorithm \cite{waldmann13c}. Based on blind-deconvolution using Independent Component Analysis \cite{waldmann12,hyvarinen00,icabook2}, we estimate the common non-Gaussian time and wavelength correlated signals and construct a systematic noise model which is then used to correct each individual time series. The advantage of these types of de-trending algorithms over others such as Gaussian Processes \cite{rasmussen06} are their non-parametric nature. This guarantees a high degree of objectivity in the de-trending as well as a simple implementation into existing code (due to the lack of parameterisation required).

\section{Lightcurve modelling}

%\begin{figure}
%  \subfloat[Single Mandel \& Agol (2002) eclipse model. The discontinuous blue line marks the out of transit baseline. The discontinuous green line marks the in-transit flux and $\delta$ defines the transit depth. Discontinuous red lines note the contact points $t_{1-4}$. ]{\includegraphics[width=8cm]{lc_note.pdf}}~~              
%  \subfloat[Schematic outline of EChO observations illustrating the changing baseline flux levels. Here blue curves illustrate the stellar out of transit spectra and the green curve the in-transit spectrum of the star. In the case of a secondary eclipse, the green curve represents the stellar spectrum only whilst the blue curve is star+planetary flux. ]{\includegraphics[width=8cm]{transspec.pdf}}
%  \caption{ Lightcurve models}
%  \label{lightcurve}
%\end{figure}

\begin{figure}[h]
\centering
\includegraphics[width=0.7\textwidth]{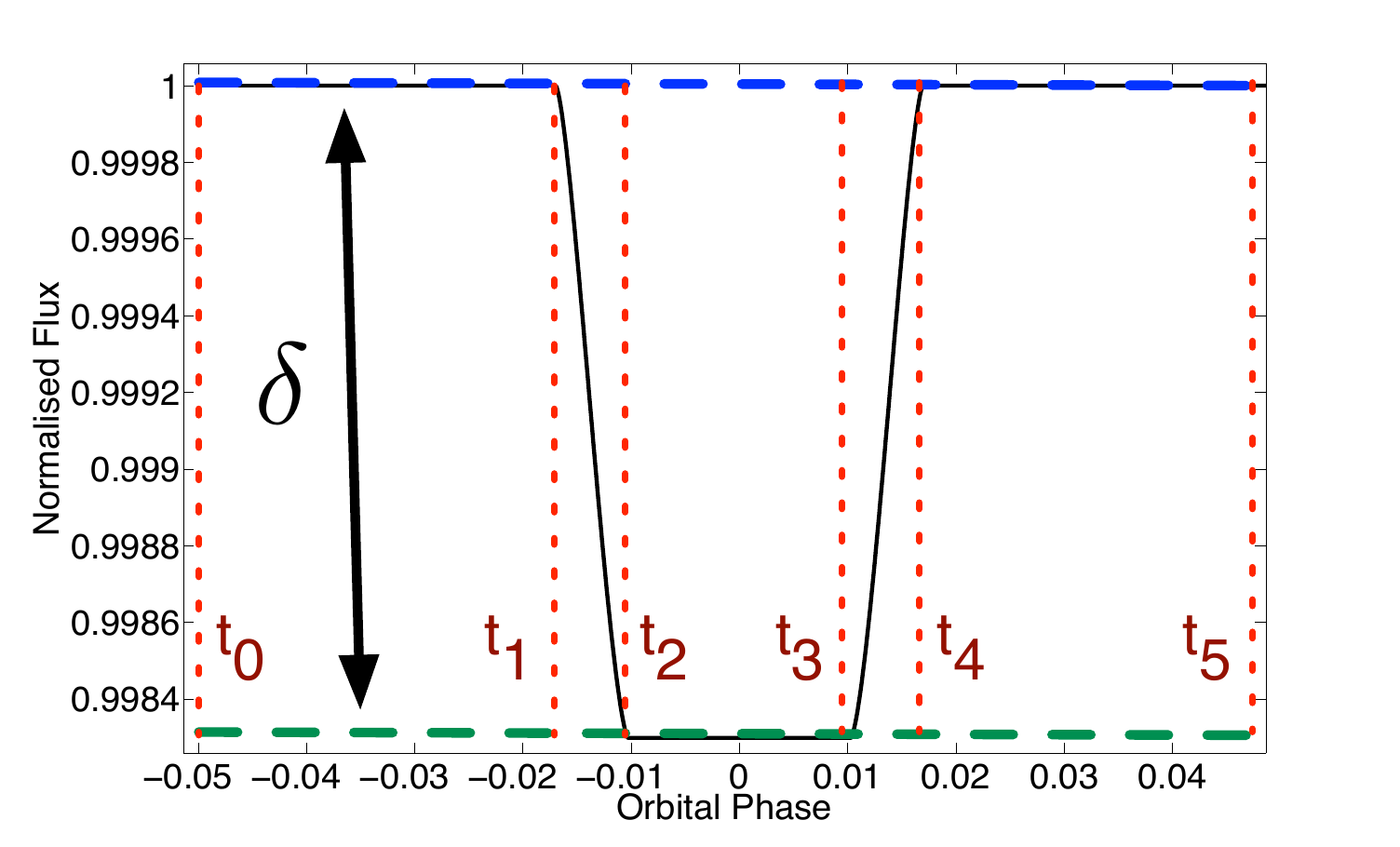}
\caption{Single Mandel \& Agol (2002) eclipse model. The discontinuous blue line marks the out of transit baseline. The discontinuous green line marks the in-transit flux and $\delta$ defines the transit depth. Discontinuous red lines note the contact points $t_{1-4}$. \label{fig:lightcurve}}
\end{figure}

Once the data is reduced and de-correlated, the pipeline provides several means of model fitting the  resulting lightcurves. The modelling is divided into two main modes: Radiomentric and Dynamic. In the simplest model assumption, the radiometric case, we simply calculate the error bar from the out-of-transit (OOT) scatter of the time series and estimate the transit depth by taking the ratio of in-transit (IT) and OOT data. 
For the Dynamic case we use a full transiting planet model \cite{mandel02} and iteratively fit for the transit depth parameter using a simplex-downhill algorithm as well as a Markov Chain Monte Carlo (MCMC) routine.

%\begin{equation}
%\mathcal{L}(\delta, {\bf d}) = \frac{1}{\sigma \sqrt{2 \pi}} \text{exp} \left [ - \frac{1}{2} \sum_{i=0}^{i=N} \left ( \frac{d_{i} - m_{i}(\delta)}{\sigma} \right )^{2} \right ]
%\end{equation}

\subsection{{\it Radiometric} data analysis}

For most cases, and for the sake of computational efficiency, the simplistic radiometric model results are desired for \echosim~observations. 
Let us assume a secondary eclipse measurement of an exoplanet. In the simplest radiometric case, we calculate the transit depth via the simple relation

\begin{equation}
\delta = F_{out} - F_{in}
\label{equ:delta}
\end{equation}

\noindent where $\delta$ is the transit depth, $F_{out}$ is the baseline flux (blue line in figure~\ref{fig:lightcurve}) and is defined as 

\begin{equation}
F_{out}~=~\frac{1}{N(t_{0-1},t_{4-5})} \left (\sum_{t=t_{0}}^{t_{1}} F_{t} + \sum_{t=t_{4}}^{t_{5}} F_{t} \right )
\end{equation}

\noindent where $t$ is the time index, $N$ the number of observations in time, $t_{0-1}$ defines pre-ingress baseline time and $t_{4-5}$ post-egress timeline (see figure~\ref{fig:lightcurve}). Similarly we define the in-transit flux as

\begin{equation}
F_{in} = \frac{1}{N(t_{2-3})} \sum_{t=t_{2}}^{t_{3}} F_{t}
\label{eq:fmin}
\end{equation}

\noindent Equation~\ref{eq:fmin} is valid for the secondary eclipse case and mid-IR transit cases where limb-darkening is negligible. To avoid the effect of limb-darkening in the case of primary eclipses in the near-IR, we borrow the `correct' transit depths from \echosim's auxiliary output files. Note that this is a valid procedure since we are dealing with an over simplistic model here. The dynamic model fitting does not assume auxiliary data. Given equation~\ref{equ:delta}, we calculate the error on $\delta$ as the sum of squares of the time series error

\begin{equation}
\sigma_{total}/\sqrt{N} = \sqrt{\sigma_{out}^{2} / N_{out}+ \sigma_{in}^{2} /N_{in}} = \frac{\sqrt{2}\sigma}{\sqrt{N}}
\label{equ:raderr}
\end{equation}

\noindent where $N$ is the number of observations and we assume that $N_{out} = N_{in} = 2N$ as well as $\sigma_{out} = \sigma_{in}$.

\subsubsection{Interpretation of {\it radiometric} model}
\label{sec:radinterp}

The assumption $\sigma_{out} = \sigma_{in}$ seems straight forward as one expects the photometric stability not to vary significantly between out-of-eclipse and in-eclipse times. The radiometric error as in equation~\ref{equ:raderr} is the correct error treatment for the observation of a single lightcurve at a single wavelength with equal lengths of out-of-transit and in-transit measurements. It assumes that no additional knowledge of the baseline (out-of-transit) flux is available and describes the state of largest ignorance, i.e. $\sigma \rightarrow \sqrt{2} \sigma$. Should additional knowledge of the baseline flux be available (via the calibration of the wavelength dependent stellar spectrum), we can reduce the normalisation error on the baseline.  
Hence for a perfect knowledge of the baseline flux level $\sigma_{total} \rightarrow \sigma$.\\

%Figure~\ref{lightcurve}b shows a schematic of the EChO observation of a single eclipse event. Here the black curves are the planetary lightcurves, the blue curves the star+planet spectra for given time stamp $i$ and the green curve the stellar spectrum only (between $t_{2} - t_{3}$). What can instantly be seen is that the baseline fluxes are correlated through wavelengths by the stellar spectrum. We can hence calibrate the baseline flux of each individual lightcurve using the stellar spectrum across wavelengths. This additional knowledge is not included in the calculation of the radiometric model error.

%\begin{figure}[H]
%\center
%\includegraphics[width=15cm]{lightcurve.pdf}
%\caption{Ble ble}
%\label{lightcurve2}
%\end{figure}

%\begin{figure}[H]
%\center
%\includegraphics[width=8cm]{intro-transspec.pdf}
%\caption{Ble ble}
%\label{lightcurve2}
%\end{figure}

\subsection{{\it Dynamic} data analysis}

\begin{figure}[h]
\centering
\includegraphics[width=8cm]{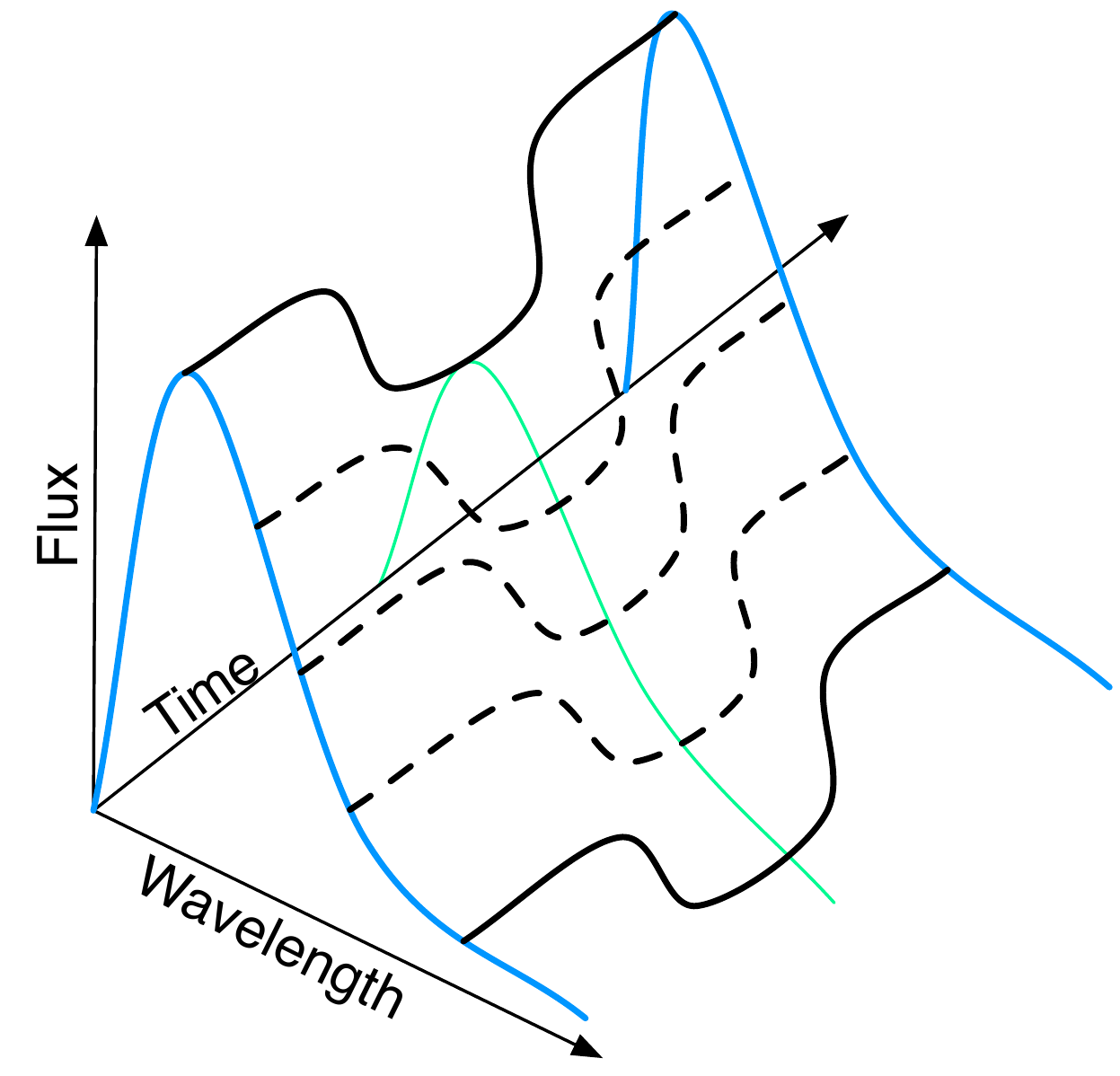}
\caption{Schematic outline of EChO observations illustrating the changing baseline flux levels. Here blue curves illustrate the stellar out of transit spectra and the green curve the in-transit spectrum of the star. In the case of a secondary eclipse, the green curve represents the stellar spectrum only whilst the blue curve is star+planetary flux.\label{fig:stellar}}
\end{figure}

\begin{figure}
  \centering
    \includegraphics[width=0.7\textwidth]{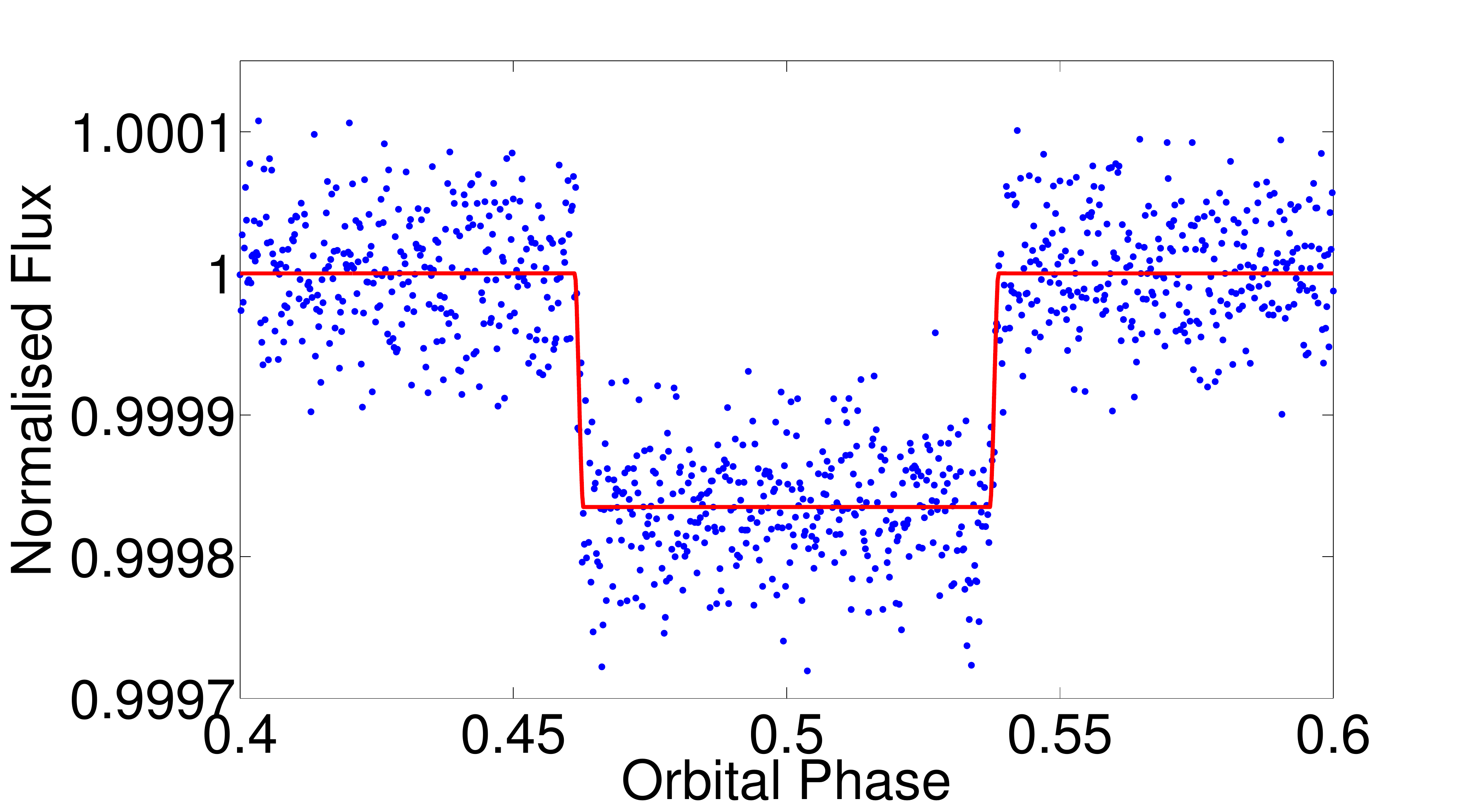}
   \caption{Normalised lightcurve of secondary eclipse of 55 Cnc e (5 eclipses co-added). Red line: analytic lightcurve model \cite{mandel02} with the eclipse depth $\delta$ as only free parameter. Note the lack of stellar limb-darkening in secondary eclipses and hence a very discrete ingress and egress.   \label{fig:lc}}
\end{figure}

\begin{figure}
  \centering
    \includegraphics[width=0.7\textwidth]{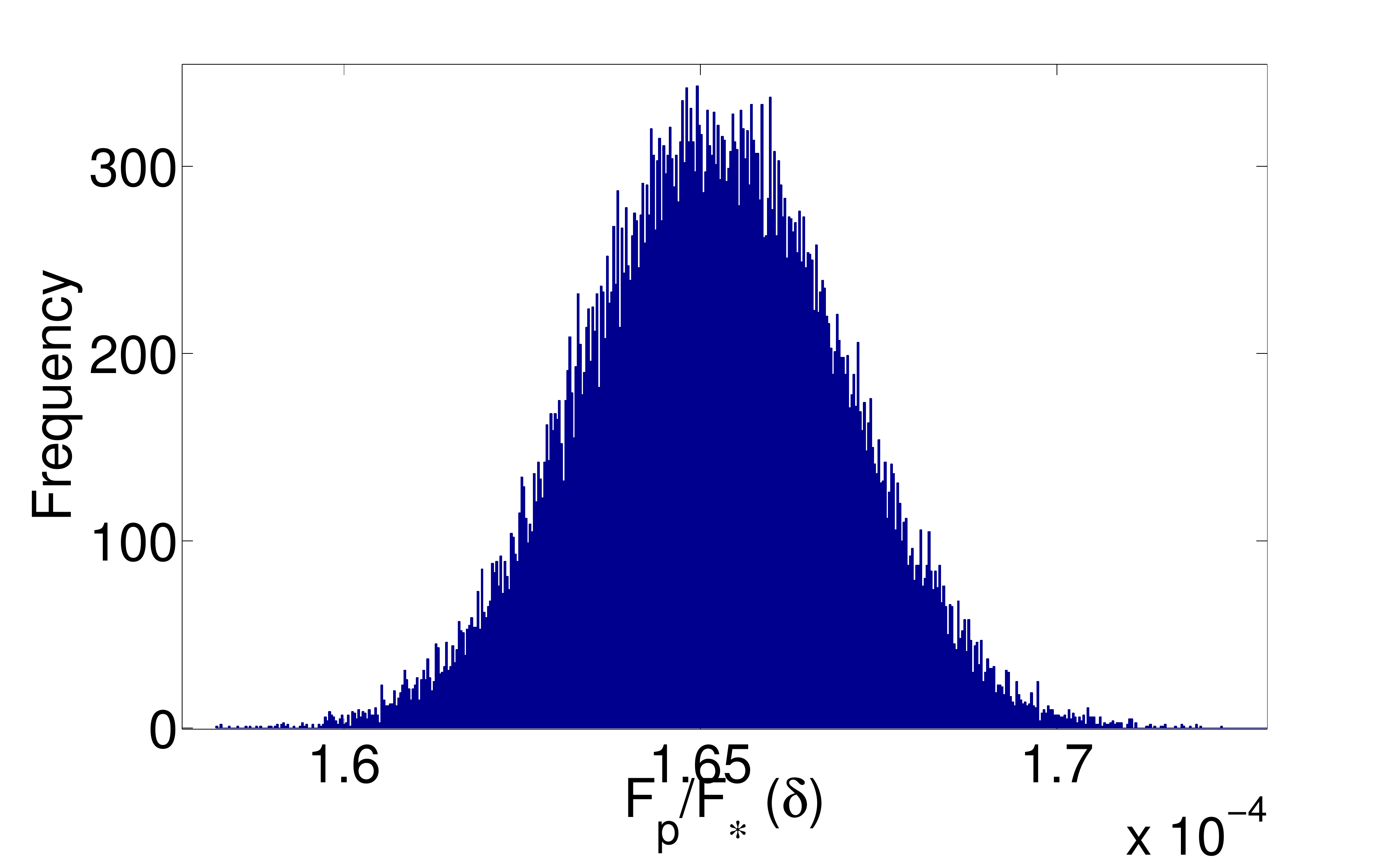}
   \caption{Histogram of MCMC chain run for 50,000 iterations. The histogram approximates the posterior distribution of the transit depth parameter $\delta$ for the model fit shown in figure~\ref{fig:lc}. \label{fig:mcmc}}
\end{figure}

Going beyond the radiometric model assumptions, \echodp~has two additional time-resolved lightcurve model modes: 1) Simplex and 2) MCMC. 

In the simplex case, we fit an analytical lightcurve model \cite{mandel02} to each individual lightcurve in wavelength space, $\lambda$. It fully supports eccentric orbit calculations following \cite{kipping08} and allows all model parameters to be fitted. For lightcurves in wavelength ranges below 5$\mu$m we assume stellar limb-darkening for primary eclipses. Here we linearly interpolate the quadratic limb-darkening coefficients of \cite{claret00} or read the limb-darkening coefficient grid provided by \echosim~to provide an exact match.  For the model minimisation we use a simplex-downhill algorithm \cite{nelder65,press07}. In this simple minimisation scheme, we obtain the error bar on the model fit using equation~\ref{equ:raderr}. Each modelling run creates a new model-fitting object in the data pipeline which allows multiple model runs (radiometric as well as dynamic) to be executed in the same instance of the \echodp. \\

We furthermore include a more computationally intensive Markov Chain Monte Carlo routine in \echodp. This routine allows us to investigate more complex scenarios and potential prior dependence (should prior knowledge on the exoplanetary or stellar spectrum be known). The posterior on the model parameter $\theta$ can be written as

%Starting from the standard bayesian argument:
%
%\begin{equation}
%p(\hat{\theta} | F, {\it H}) =  \frac{p(F | \theta, {\it H}) p (\theta | {\it H})}{p (F | {\it H})}
%\end{equation}
%
%\noindent where $p(\hat{\theta} | F, {\it H}) $ is the posterior of the parameters given the flux data $F$ (and the model hypothesis $H$), $\hat{\theta}$ is the estimated parameters, $p(F | \theta, {\it H})$ is the likelihood function, also written as $\mathcal{L}(\theta)$, $p (F | {\it H})$ is the prior knowledge on the system (written as $\pi(\theta)$), $p (F | {\it H})$ is the Bayesian evidence and noted as $\mathcal{Z}$:
%
%\begin{equation}
%p(\hat{\theta} | F) =  \frac{\mathcal{L}(\theta) \pi(\theta)}{\mathcal{Z}}
%\end{equation}
%
%\noindent given that $\mathcal{Z} = \int \mathcal{L} (\theta) \pi (\theta) d^{n} \theta$ only depends on the parameters $\theta$ and not on the data, we do not need to calculate it to obtain the model-fit posteriors. We hence have

%\begin{equation}
%\mathcal{Z} = \int \mathcal{L} (\theta) \pi (\theta) d^{n} \theta
%\end{equation}

\begin{equation}
p(\hat{\theta} | F) \propto \mathcal{L}(\theta) \pi(\theta).
\end{equation}

\noindent where $\mathcal{L}(\theta)$ is the model likelihood and $\pi(\theta)$ the prior distribution on the parameter $\theta$. The likelihood is here assumed to be Gaussian and is given by

\begin{equation}
\mathcal{L}(\theta, {\bf d}) = \frac{1}{\sigma \sqrt{2 \pi}} \text{exp} \left [ - \frac{1}{2} \sum_{t=0}^{t=N} \left ( \frac{d_{t} - \Phi_{t}(\theta)}{\sigma} \right )^{2} \right ]
\end{equation}

\noindent where ${\bf d}$ is the data column vector, and $d_{t}$ and $\Phi_{t}(\theta)$ are the datum and lightcurve model at given timestamp $t$. 

We use the PyMC\footnote{$https://github.com/pymc-devs/pymc$} package implementing the adaptive Metropolis Hastings algorithm of \cite{haario06}. The MCMC chains are typically run with 20,000 iterations taking the minimised result of the simplex-downhill algorithm as starting value to minimise burn-in time \cite{brooks11} which we restrict to 1000 iterations. We here present the univariate version of the likelihood as in most cases all transit parameters but the depth, $\delta$, are fixed. To minimise parameter covariances for multiple free parameters one can follow parameterisation by \cite{bakos08} or \cite{burke07}. 
Using a bayesian approach, we can investigate more complex model solutions such as the impact of the stellar variability on the normalisation of individual lightcurves. Figure~\ref{fig:stellar} illustrates a time series observation of a transiting exoplanet over a wide range of wavelengths. Here the blue curves represent the stellar spectrum, the black curves the time dependent flux variation due to the transiting extrasolar planet with the green line marking the minimum flux. As discussed in section~\ref{sec:radinterp}, if all time series measurements are assumed to be independent of each other (i.e. not correlated in wavelength), we must assume an error of $\sqrt{2}\sigma$ on the measurement, given the uncertainty of the OOT normalisation. However, it is clear from figure~\ref{fig:stellar} that OOT flux of individual time series is correlated in $\lambda$ through the stellar spectrum.  For a perfect correlation (i.e. absolute knowledge on the correct normalisation of the individual time series) the measurement error hence reduces to $\sigma$. Hence the normalisation error, $\sigma_{norm}$, is bound by $ 0 \leq \sigma_{norm} \leq \sqrt{2}$. 

 We can now express the likelihood of our observation, $\mathcal{L}$, as product of the likelihood of the lightcurve model, $\mathcal{L}(\theta)$ and the stellar spectrum model $\mathcal{L}(\varphi)$. Note that by taking the product we implicitly assume statistical independence between lightcurve and stellar spectra models and below we explicitly assume a Gaussian noise model

\begin{equation}
\mathcal{L} = \mathcal{L}(\theta) \mathcal{L}(\varphi) =  e^{-\frac{1}{2} \chi^{2}(\theta)} ~e^{-\frac{1}{2} \chi^{2}(\varphi)}
\label{equ:likelihood}
\end{equation}

\noindent where $\chi^{2}$ is the chi-squared distribution. We can now write the log-likelihood as follows
%\noindent where the chi-squared distributions of both lightcurve and stellar models are given by $\chi_{i}^{2} = \sum^{N}_{i=1} \left ( \frac{ F_{i,j} - \Phi_{i}}{\sigma_{i}} \right )^{2}$ and $\chi_{j}^{2} = \sum^{M}_{j=1} \left ( \frac{ \bar{F}_{i=t_{2-3},j} - \Psi_{j}}{\sigma_{j}} \right )^{2}$

\begin{align}
\text{log} \mathcal{L} &= -\frac{1}{2} \sum^{N}_{t=1} \left ( \frac{ F_{t,\lambda} - \Phi(\theta_{t})}{\sigma_{t}} \right )^{2} \\\nonumber
&-\frac{1}{2} \sum^{M}_{\lambda=1} \left ( \frac{ \bar{F}_{t=t_{2-3},\lambda} - \Psi(\theta_{\lambda})}{\sigma_{\lambda}} \right )^{2}
\end{align}

\noindent where $\Phi(\theta_{t})$ is the lightcurve model for given time index $t$, $\Psi(\theta_{\lambda})$ is the stellar model for given wavelength index $\lambda$, $M$ is the number of resolution elements in the spectrum and $\sigma_{t}$ and $\sigma_{\lambda}$ are the flux uncertainties on the time series and the stellar spectrum respectively. Note that these error terms are not equivalent and also note that $\bar{F}_{t=t_{2-3},\lambda}$ is the averaged stellar spectrum from time interval $t_{2} - t_{3}$. \\

\section{Outputs}

\begin{figure}
  \centering
    \includegraphics[width=0.85\textwidth]{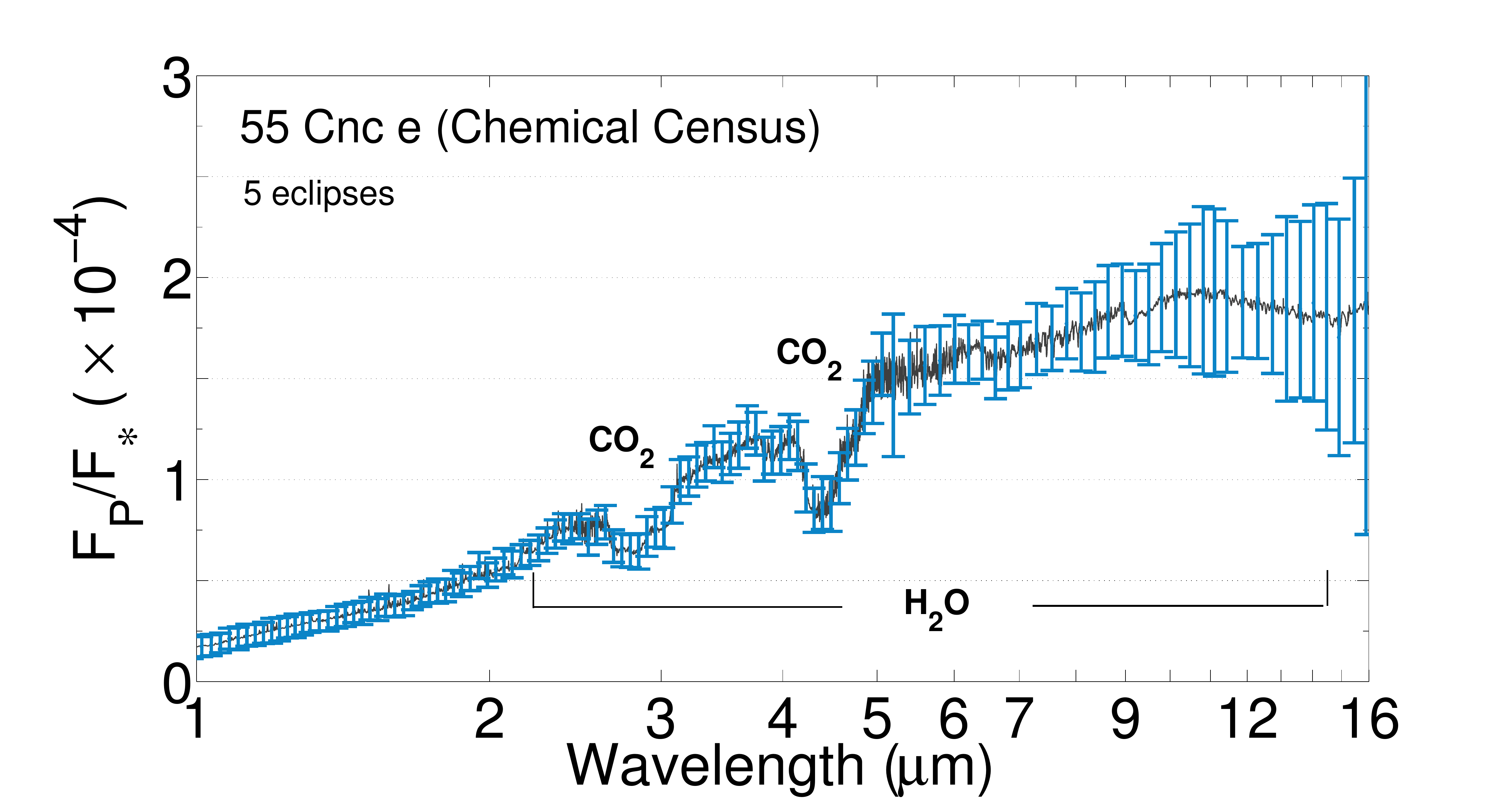}
   \caption{Final spectrum generated from \echodp~outputs for 55 Cnc e secondary eclipse run in Chemical census mode (i.e. 5 eclipses stacked, R = 50 for $\lambda < 5 \mu$m and R = 30 for $\lambda > 5 \mu$m). Blue: error bars derived from \echodp. Grey: planetary emission spectrum read into \echosim. We marked prominent emission/absorption features. \label{fig:result}}
\end{figure}

\begin{figure}
  \centering
    \includegraphics[width=0.85\textwidth]{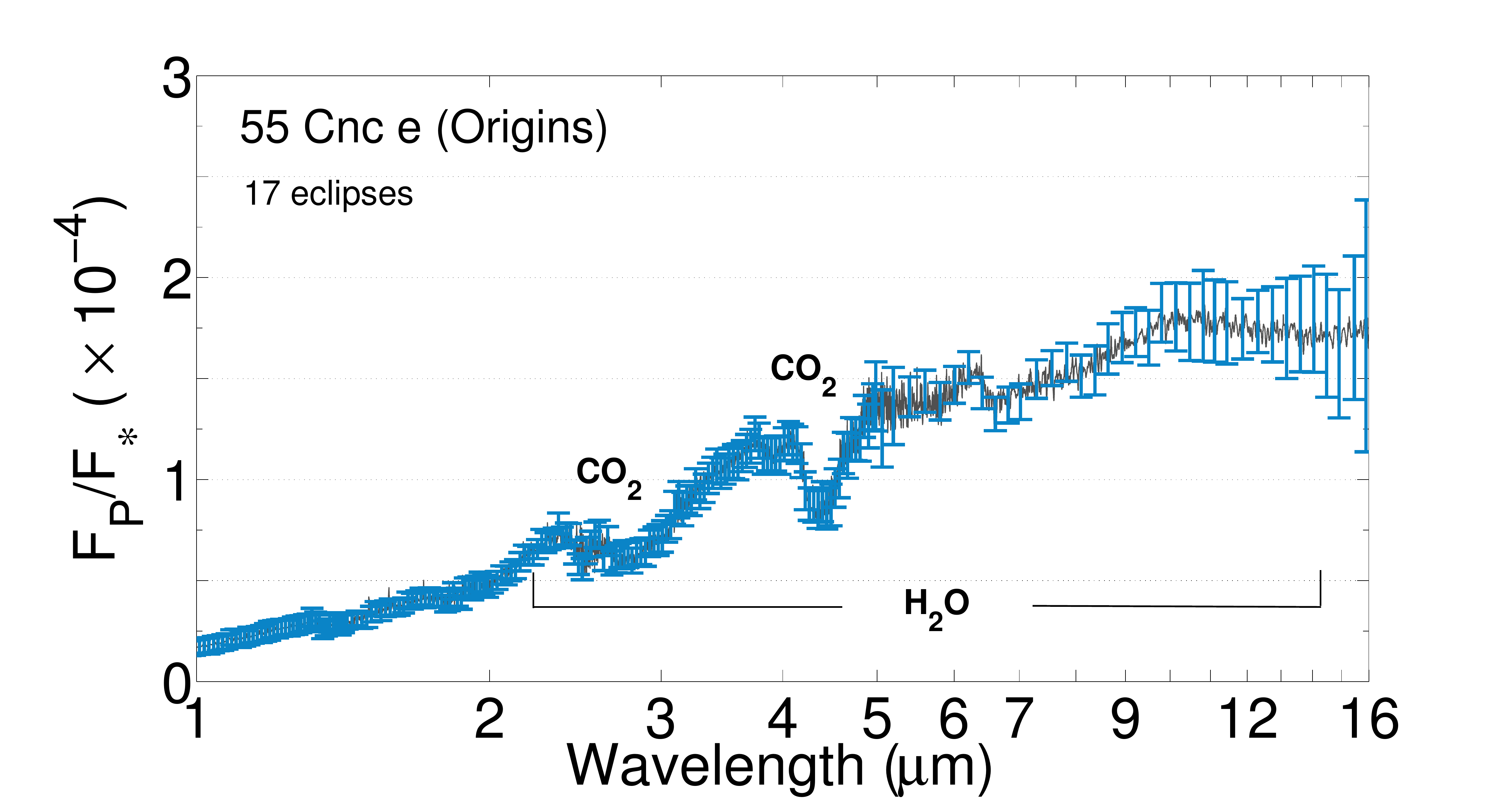}
   \caption{Final spectrum generated from \echodp~outputs for 55 Cnc e secondary eclipse run in Origin  mode (i.e. 17 eclipses stacked, R = 100 for $\lambda < 5 \mu$m and R = 30 for $\lambda > 5 \mu$m). Blue: error bars derived from \echodp. Grey: planetary emission spectrum read into \echosim. We marked prominent emission/absorption features. \label{fig:result2}}
\end{figure}

\begin{figure}
  \centering
    \includegraphics[width=0.85\textwidth]{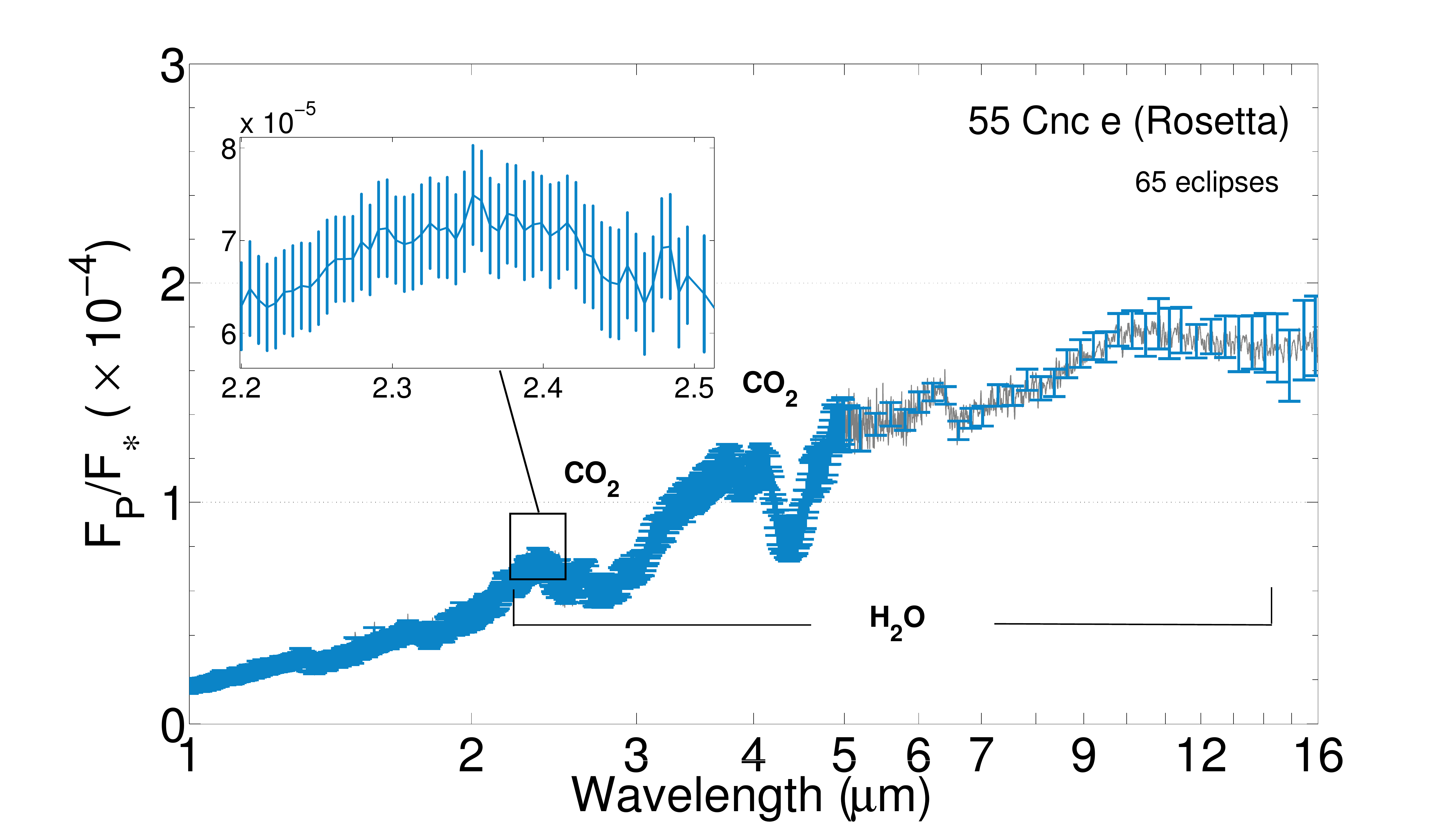}
   \caption{Final spectrum generated from \echodp~outputs for 55 Cnc e secondary eclipse run in Rossetta mode (i.e. 65 eclipses stacked, R = 300 for $\lambda < 5 \mu$m and R = 30 for $\lambda > 5 \mu$m). Blue: error bars derived from \echodp. Grey: planetary emission spectrum read into \echosim. Inset is a zoom into the 2.2 - 2.5 $\mu$m wavelength region. We marked prominent emission/absorption features. \label{fig:result3}}
\end{figure}

Two types of outputs are provided: spectra in ascii format and python-pickel\footnote{$http://docs.python.org/2/library/pickle.html$} objects. For each individual lightcurve fitting, \echodp~provides an ascii file containing wavelength, measured flux and error. The pickle file contains all parameters, intermediate and final data products allowing for an exact reproducibly of results. Figure~\ref{fig:result} shows the final spectrum for 55 Cnc e in the Chemical Census mode (blue error bars). Figures~\ref{fig:result2} and \ref{fig:result3} show the same simulation for the Origins and Rosetta stone observing modes of \echo.

\section{Discussion \& Conclusion}

\echodp~is a custom built data reduction and analysis pipeline for the \echosim~end-to-end mission simulator of the \echo~mission concept. 

Despite its customised nature, we have developed the pipeline with easy adaptability (through its fully object-orientated programming ) to other instruments and data-sets in mind. The pipeline features state of the art data de-correlation algorithms as well as a full Bayesian analysis implementation via adaptive MCMC. Both these aspects, the de-trending as well as the exploration of stellar variability are not required for the current version of \echosim~(version 3.x) but included with future releases. These releases will have special emphasis on realistic stellar noise simulations \cite{ballerini12} as well as more advanced non-Gaussian instrument systematics.

%\begin{acknowledgements}
%\end{acknowledgements}

\bibliographystyle{spphys}       % APS-like style for physics
\bibliography{echo-obspipeline-paper}   % name your BibTeX data base

%\begin{thebibliography}{}
%\end{thebibliography}

\end{document}